\documentclass[twocolumn,english,superscriptaddress,showpacs,prl]{revtex4}
\usepackage{tgtermes}
\usepackage[utf8]{inputenc}
\usepackage{color}
\usepackage{babel}
\usepackage{bm}
\usepackage{graphicx}
\usepackage{esint}
\usepackage[unicode=true,
 bookmarks=false,
 breaklinks=false,pdfborder={0 0 1},backref=false,colorlinks=true]
 {hyperref}
\hypersetup{
 citecolor=blue}

\makeatletter
\@ifundefined{textcolor}{}
{%
 \definecolor{BLACK}{gray}{0}
 \definecolor{WHITE}{gray}{1}
 \definecolor{RED}{rgb}{1,0,0}
 \definecolor{GREEN}{rgb}{0,1,0}
 \definecolor{BLUE}{rgb}{0,0,1}
 \definecolor{CYAN}{cmyk}{1,0,0,0}
 \definecolor{MAGENTA}{cmyk}{0,1,0,0}
 \definecolor{YELLOW}{cmyk}{0,0,1,0}
}

\usepackage{babel}

\makeatother

\begin{document}

\title{Topological phases with long-range interactions}

\author{Z.-X. Gong}
\email{gzx@umd.edu}

\affiliation{Joint Quantum Institute, NIST/University of Maryland, College Park,
Maryland 20742, USA}

\affiliation{Joint Center for Quantum Information and Computer Science, NIST/University
of Maryland, College Park, Maryland 20742, USA}

\author{M. F. Maghrebi}

\affiliation{Joint Quantum Institute, NIST/University of Maryland, College Park,
Maryland 20742, USA}

\affiliation{Joint Center for Quantum Information and Computer Science, NIST/University
of Maryland, College Park, Maryland 20742, USA}

\author{A. Hu}

\affiliation{Joint Quantum Institute, NIST/University of Maryland, College Park,
Maryland 20742, USA}

\affiliation{Department of Physics, American University, Washington, DC 20016,
USA}

\author{M. L. Wall}

\affiliation{JILA, NIST/University of Colorado, Boulder, CO 80309, USA}

\author{M. Foss-Feig}

\affiliation{Joint Quantum Institute, NIST/University of Maryland, College Park,
Maryland 20742, USA}

\affiliation{Joint Center for Quantum Information and Computer Science, NIST/University
of Maryland, College Park, Maryland 20742, USA}

\author{A. V. Gorshkov}

\affiliation{Joint Quantum Institute, NIST/University of Maryland, College Park,
Maryland 20742, USA}

\affiliation{Joint Center for Quantum Information and Computer Science, NIST/University
of Maryland, College Park, Maryland 20742, USA}
\begin{abstract}
Topological phases of matter are primarily studied in systems with
short-range interactions. In nature, however, non-relativistic quantum
systems often exhibit long-range interactions. Under what conditions
topological phases survive such interactions, and how they are modified
when they do, is largely unknown. By studying the symmetry-protected
topological phase of an antiferromagnetic spin-1 chain with $1/r^{\alpha}$
interactions, we show that two very different outcomes are possible,
depending on whether or not the interactions are frustrated. While
non-frustrated long-range interactions can destroy the topological
phase for $\alpha\lesssim3$, the topological phase survives frustrated
interactions for all $\alpha>0$. Our conclusions are based on strikingly
consistent results from large-scale matrix-product-state simulations
and effective-field-theory calculations, and we expect them to hold
for more general interacting spin systems. The models we study can
be naturally realized in trapped-ion quantum simulators, opening the
prospect for experimental investigation of the issues confronted here. 
\end{abstract}

\pacs{75.10.Jm, 75.10.Pq, 03.65.Vf}

\maketitle
Since the discovery of topological insulators \cite{hasan_colloquium:_2010,qi_topological_2011,moore_birth_2010},
there has been tremendous interest in exploring various topological
phases of matter, both theoretically \cite{wen_quantum_2007,chen_symmetry-protected_2012}
and experimentally \cite{hsieh_topological_2008,hafezi_imaging_2013,jotzu_experimental_2014}.
Topological phases are generally associated with—and derive much of
their presumed utility from—stability against \emph{local} perturbations.
But precisely what constitutes ``local'' in this context is a subtle
issue; power-law decaying ($1/r^{\alpha}$) interactions, which are
present in many experimental systems, do not necessarily qualify \cite{eisert_breakdown_2013,hauke_spread_2013,metivier_spreading_2014}.
Recent theoretical advances have begun to elucidate the conditions
under which long-range interacting systems maintain some degree of
locality \cite{gong_persistence_2014,foss-feig_nearly_2015}, potentially
providing some insight into effects of long-range interactions on
topological phases of matter. And recently, explicit theoretical evidence
of topological order has been found in a variety of long-range interacting
systems, including dipolar spins \cite{manmana_topological_2013}
or bosons \cite{dalmonte_homogeneous_2011}, fermions with long-range
pairing \cite{vodola_kitaev_2014} and hopping \cite{pientka_unconventional_2014,vodola_longrange_2015},
and electrons with Coulomb interactions \cite{hohenadler_phase_2014}.
These results notwithstanding, a complete understanding of how topological
phases respond to the addition of long-range interactions is still
lacking.

The stability of topological phases to small local perturbations is
intimately connected to the existence of a bulk excitation gap \cite{bravyi_topological_2010,ft1},
and the introduction of long-range interactions to a short-range Hamiltonian
supporting a topological phase poses several potential challenges
to this connection. First, even if the gap remains finite, long-range
interactions can change the ground state correlation decay from exponential
to power-law \cite{hastings_spectral_2006,schachenmayer_dynamical_2010,vodola_kitaev_2014,vodola_longrange_2015}.
Thus topological phases with local interactions are, at the very least,
subject to qualitative changes in their long-distance correlations.
Second, the gap can in principle close in the presence of long-range
interactions, even when they decay fast enough that the total interaction
energy remains extensive \cite{bravyi_topological_2010,michalakis_stability_2013}.
Third, long-range interactions have the ability to change the effective
dimensionality of the system \cite{fisher_critical_1972,bruno_absence_2001},
and thus might change the topological properties even if the gap does
not close \cite{vodola_kitaev_2014,vodola_longrange_2015}. We emphasize
that the understanding of these issues is not of strictly theoretical
interest. Many of the promising experimental systems for exploring
or exploiting topological phases of matter, e.g.\ dipolar molecules
\cite{yan_observation_2013,hazzard_many-body_2014,peter_anomalous_2012},
magnetic \cite{de_paz_nonequilibrium_2013} or Rydberg atoms \cite{schaus_observation_2012},
trapped ions \cite{deng_effective_2005,hauke_complete_2010,koffel_entanglement_2012,britton_engineered_2012,richerme_non-local_2014,jurcevic_quasiparticle_2014},
and atoms coupled to multi-mode cavities \cite{douglas_quantum_2015},
are accurately described as quantum lattice models with power-law
decaying interactions. The unique controllability and measurement
precision afforded by these systems hold great promise to improve
our understanding of topological phases \cite{yao_topological_2012,yao_realizing_2013,zhu_probing_2011,wang_probe_2014},
but first we must reliably determine when—despite their long-range
interactions—they can be expected to harbor the topological phases
that have been theoretically explored for short-range interacting
systems.

To address these general questions, in this manuscript we study a
spin-1 chain with antiferromagnetic Heisenberg interactions, which
is a paradigmatic model exhibiting a symmetry-protected topological
(SPT) phase \cite{chen_symmetry_2013,pollmann_symmetry_2012}. Specifically,
we consider two extensions of the short-range version of this model
by including long-range interactions that decay either as $\mathcal{J}_{\alpha}(r)=1/r^{\alpha}$
or as $\mathcal{J}_{\alpha}^{\prime}(r)=(-1)^{r-1}/r^{\alpha}$, which
could be simulated in trapped-ion based experiments for $0<\alpha<3$
\cite{cohen_proposal_2014,senko_experimental_2014}. Based on a combination
of large-scale variational matrix-product-state (MPS) simulations
and field-theory calculations, we establish and explain a number of
important and potentially general consequences of long-range interactions.
The $\mathcal{J}_{\alpha}^{\prime}(r)$ interactions are unfrustrated,
being antiferromagnetic (ferromagnetic) between spins on the opposite
(same) sub-lattice. In this case, numerics and field theoretic arguments
suggest the destruction of the topological phase for $\alpha\lesssim3$,
accompanied by a closing of the bulk excitation gap and spontaneous
breaking of a continuous symmetry in 1$D$, consistent with other
recent findings on the relevance of long-range interactions for $\alpha<D+2$
in $D$-dimensional quantum systems \cite{maghrebi_causality_2015,maghrebi_continuous_2015}.
The $\mathcal{J}_{\alpha}(r)$ interactions are frustrated, and, remarkably,
do not close the bulk excitation gap for any $\alpha>0$. In addition,
two key properties of the SPT phase, a doubly degenerate entanglement
spectrum \cite{pollmann_entanglement_2010-1} and a non-vanishing
string-ordered correlation \cite{kennedy_hidden_1992}, are both preserved.
However, because of the long-range interactions, spin-spin correlations
and the edge-excitation amplitudes only decay exponentially within
some intermediate distance scale, after which they decay algebraically.
We expect these qualitative changes to be quite general, occurring
in other long-range interacting systems in which the topological phase
survives.

\begin{figure}
\includegraphics[width=0.95\columnwidth]{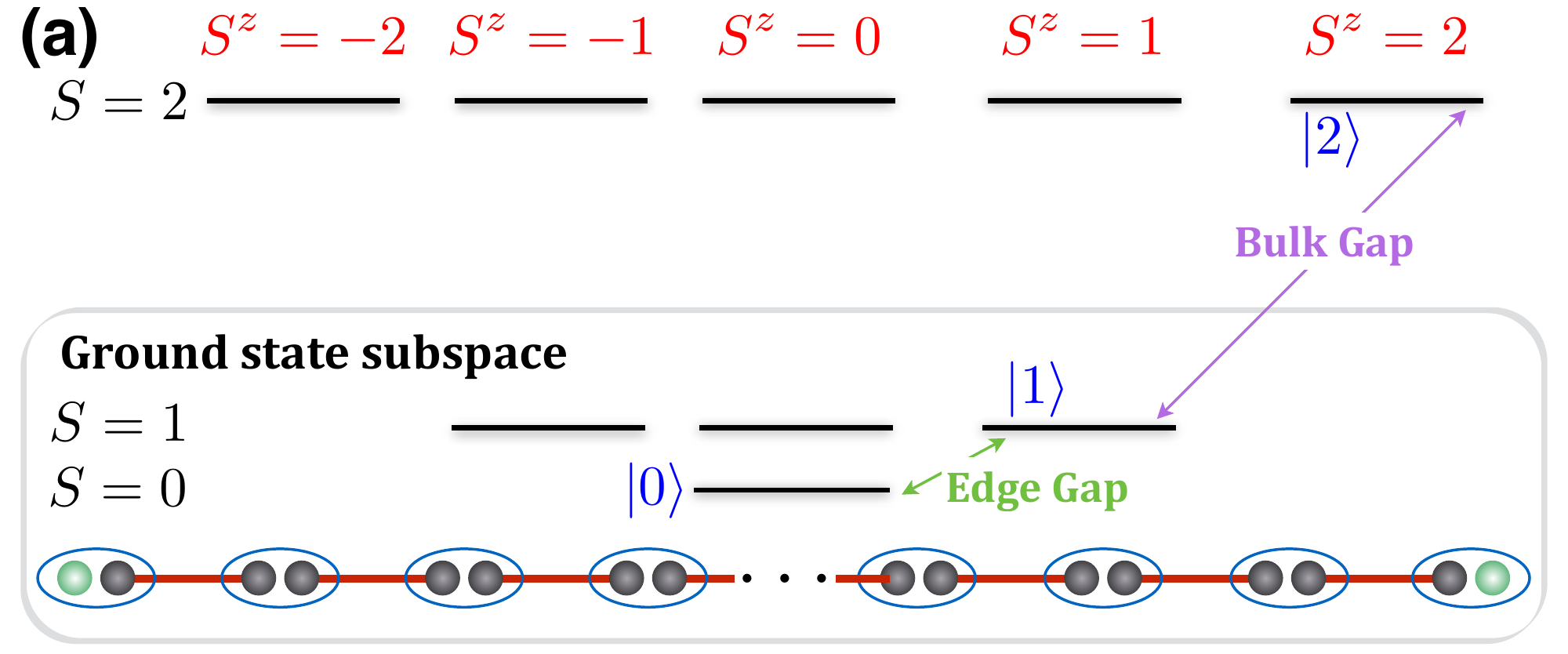}

\includegraphics[width=0.49\columnwidth]{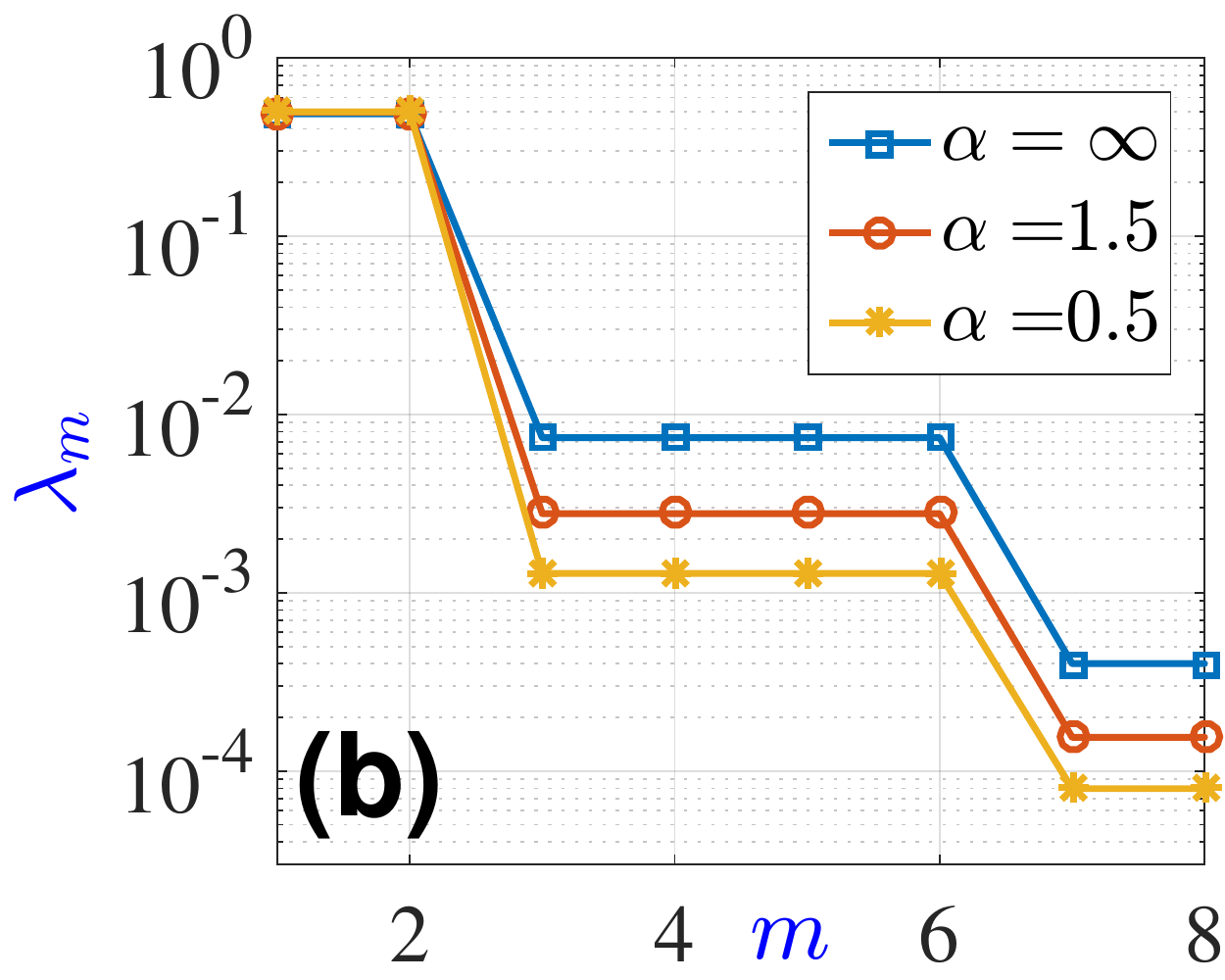}\includegraphics[width=0.49\columnwidth]{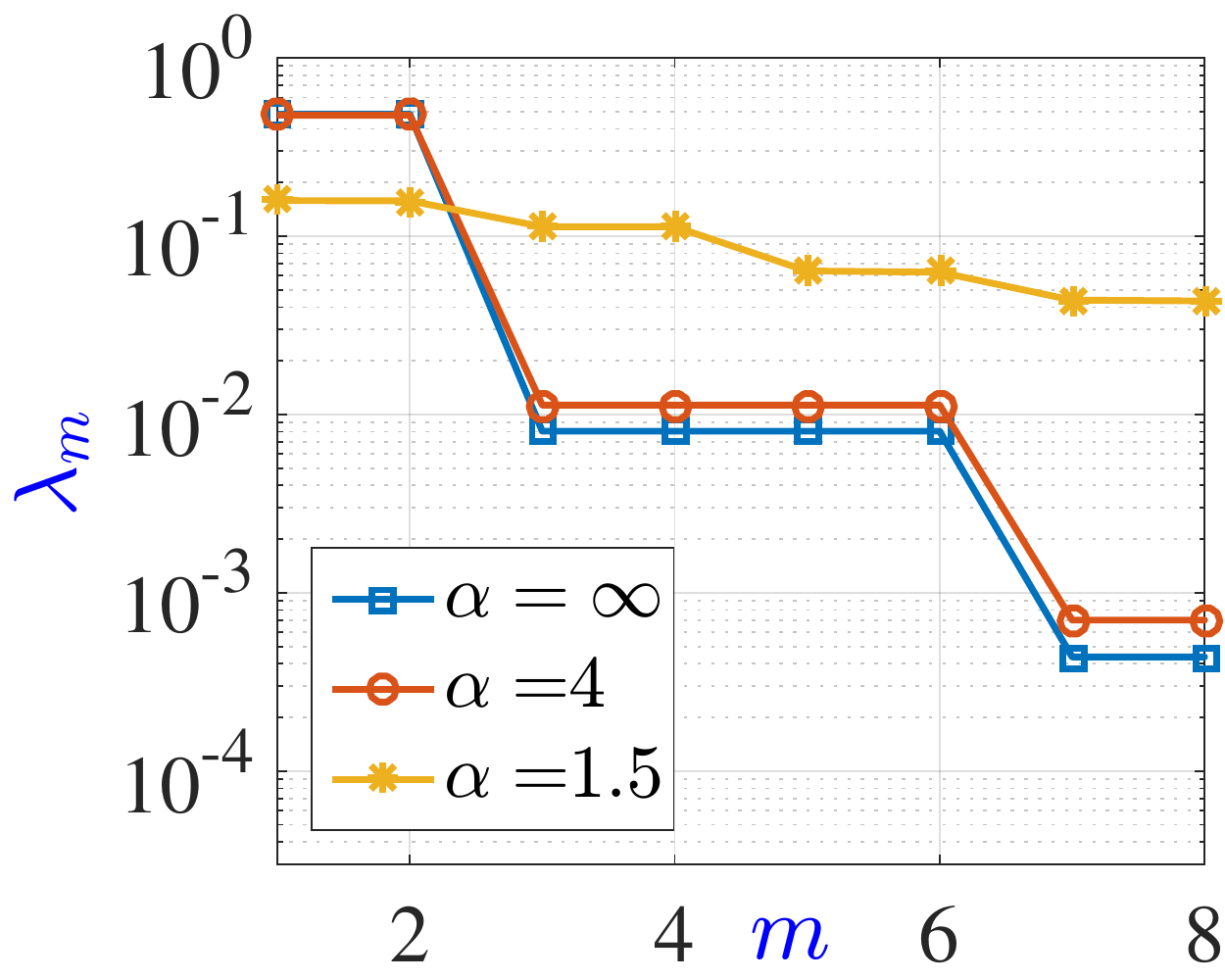}
\caption{\label{fig1}(a) Low-lying energy levels of the Haldane chain for
even $L$. The entanglement structure of ground states is shown at
the bottom. The ground states in the total $S^{z}=0,1,2$ subspace
are named $|0\rangle,|1\rangle,|2\rangle$ and have energies $E_{0},E_{1},E_{2}$.
(b-c) The $m^{th}$ largest value $\lambda_{m}$ ($m=1,2\cdots8$)
of the ground-state entanglement spectrum for $H_{\alpha}$ (b) and
$H_{\alpha}^{\prime}$ (c) using finite-size MPS calculations with
$L=200$. We choose the $|1\rangle$ state to avoid extra entanglement
between edge spins. For $H_{\alpha}^{\prime}$, the entanglement spectrum
for $1.5\le\alpha\le4$ will exhibit a smooth crossover between the
$\alpha=1.5$ and $\alpha=4$ cases due to the finite system size,
but we expect a sharp transition at some $\alpha_{c}\lesssim3$ in
the thermodynamic limit. The exact pair degeneracies in $\{\lambda_{m}\}$
are a result of the spatial-inversion symmetry protecting the topological
phase \cite{pollmann_entanglement_2010-1,pollmann_symmetry_2012}.}
\end{figure}

\emph{Model.—}We consider a spin-1 chain with either frustrated or
unfrustrated long-range Heisenberg interactions: 
\begin{equation}
H_{\alpha}=\!\!\sum_{j,r>0}\!\!\mathcal{J}_{\alpha}(r)\bm{S}_{j}\cdot\bm{S}_{j+r},\hspace{1em}H_{\alpha}^{\prime}=\!\!\sum_{j,r>0}\!\!\mathcal{J}_{\alpha}^{\prime}(r)\bm{S}_{j}\cdot\bm{S}_{j+r}.\label{eq:H}
\end{equation}
With only nearest-neighbor interactions ($\alpha\rightarrow\infty$),
$H_{\infty}=H_{\infty}^{\prime}$ is usually called the \emph{Haldane
chain}, which has been extensively studied theoretically \cite{haldane_continuum_1983,haldane_nonlinear_1983,affleck_proof_1986},
numerically \cite{white_density_1992,white_numerical_1993,sorensen_large-scale_1993,sorensen_equal-time_1994,sorensen_sk_1994},
and experimentally \cite{batista_electron_1999,yoshida_energy_2005}.
The low-lying states of the Haldane chain are shown in Fig.\,\ref{fig1}(a)
for an open boundary chain with even size $L$. The unique ground
state has total spin $S=0$. The first set of excited states has $S=1$
($\hbar=1$), contains spin excitations only near the edge of the
chain, and is separated from the ground state by an energy gap (\emph{edge
gap}) that is exponentially small in $L$ and topologically protected.
Consequently, these excited states belong to a degenerate ground-state
subspace in the thermodynamic ($L\rightarrow\infty$) limit. The second
set of excited states all have $S=2$, contain spin excitations in
the bulk of the chain, and have an energy gap (\emph{bulk gap}) that
converges to a finite value when $L\rightarrow\infty$. The entanglement
structure of the four ground states is close to that of the Affleck-Kennedy-Lieb-Tasaki
(AKLT) states \cite{affleck_rigorous_1987} shown at the bottom of
Fig.\,\ref{fig1}(a), where each spin-1 is decomposed into two spin-1/2s,
pairs of spin-1/2s on neighboring sites form singlets, and the system
is finally projected back onto the spin-1s. The four quasi-degenerate
ground states correspond to the four states formed by the two unpaired
spin-1/2s at the edge.

We use variational MPS calculations \cite{schollwock_density-matrix_2011,crosswhite_applying_2008,wall_out--equilibrium_2012,ft3},
to determine the ground-state entanglement structure of $H_{\alpha}$
and $H_{\alpha}^{\prime}$ in Fig.\,\ref{fig1}(b-c). For $\alpha>0$
($\alpha>3$), the ground-state entanglement spectrum of $H_{\alpha}$
($H_{\alpha}^{\prime}$), defined as the eigenvalues of the left/right
half-chain's reduced density matrix, is dominated by the two largest
degenerate eigenvalues $\lambda_{1}=\lambda_{2}\approx0.5$. This
can be understood heuristically as the result of cutting a spin-1/2
singlet in the AKLT state, and suggests the survival of the topological
Haldane phase. For $H_{\alpha}^{\prime}$ with $\alpha\lesssim3$,
the entanglement spectrum has an entirely different structure, and
we will study the related ground state properties below.

\emph{Effective field theory.—}The low energy physics of the Haldane
chain can be understood via field theoretic analysis due to Haldane
\cite{haldane_nonlinear_1983} and Affleck \cite{affleck_quantum_1985};
here we build on their work to provide a field theoretic treatment
of the long-range interacting model. We begin by decomposing the spin
operators into staggered and uniform fields, $\bm{n}(2i+\frac{1}{2})=(\bm{S}_{2i}-\bm{S}_{2i+1})/2$
and $\bm{l}(2i+\frac{1}{2})=(\bm{S}_{2i}+\bm{S}_{2i+1})/2$. The intuition
behind this decomposition is that the classical ground state of both
$H_{\alpha}$ and $H_{\alpha}^{\prime}$ is Néel-ordered for any $\alpha>0$,
with $\bm{n}^{2}(x)=1$ and $\bm{l}(x)=0$. We therefore expect that
in the quantum ground state $\bm{n}^{2}(x)\approx1$, while $\bm{l}(x)\approx0$
represents small quantum fluctuations in the direction of $\bm{n}(x)$.
Importantly, we expect that only long-wavelength fluctuations of $\bm{n}(x)$
and $\bm{l}(x)$ will be important at low energy. In momentum space,
we can write $H_{\alpha}\approx\int dq\big(\omega(q)|\bm{n}(q)|^{2}+\Omega(q)|\bm{l}(q)|^{2}\big)$
and $H_{\alpha}^{\prime}\approx\int dq\big(\Omega(q)|\bm{n}(q)|^{2}+\omega(q)|\bm{l}(q)|^{2}\big)$
\cite{ft2}, with 
\begin{equation}
\omega(q)=2\sum_{r=1}^{\infty}\mathcal{J}_{\alpha}^{\prime}(r)\cos qr,~~\Omega(q)=2\sum_{r=1}^{\infty}\mathcal{J}_{\alpha}(r)\cos qr.
\end{equation}

For any $\alpha>0$, $\omega(q)$ is analytic at small $q$ and can
be expanded as $\omega_{0}+\omega_{2}q^{2}+O(q^{4})$, whereas $\Omega(q)$
is non-analytic at small $q$ with an expansion $\Omega_{0}+\Omega_{2}q^{2}+\lambda|q|^{\alpha-1}+O(q^{4})$.
The coefficients $\omega_{0,2}$, $\Omega_{0,2}$, and $\lambda$
depend on $\alpha$, but their exact values are not important for
the following analysis. Physically, the analyticity (non-analyticity)
of the spectrum arises because the long-range interactions interfere
destructively (constructively) for the staggered field. Keeping only
the lowest non-trivial order in $q$ for the dispersion of both $\bm{n}(q)$
and $\bm{l}(q)$ turns out to be sufficient for obtaining qualitatively
correct behavior of the excitation gap. Therefore, we keep only the
$0^{{\rm th}}$ order term in the dispersion of $\bm{l}(q)$, and
the next-leading term in the dispersion of $\bm{n}(q)$ (for $\bm{n}(q)$,
the $0^{{\rm th}}$ order term only adds a constant to the Hamiltonian
due to the constraint $\bm{n}^{2}(x)=1$). Thus for $\alpha>0$ ($\alpha>3$)
the Hamiltonian $H_{\alpha}$ ($H_{\alpha}^{\prime}$) is approximately
given by (ignoring the order-unity coefficients) $H_{\alpha}\sim H_{\alpha}^{\prime}\sim\int dq\big(q^{2}|\bm{n}(q)|^{2}+|\bm{l}(q)|^{2}\big)$.
When the zero-temperature partition function is expressed as a coherent-spin-state
path integral, the action is quadratic in the field $\bm{l}$ and
it can be integrated out \cite{fradkin_field_2013,sachdev_quantum_2011}.
The remaining path integral over the staggered field $\bm{n}$ is
a 1+1$D$ $O(3)$ nonlinear sigma model, with Lagrangian density (nonlinear
constraint $\bm{n}^{2}(x)=1$ implied) 
\begin{equation}
\mathcal{L}(x)\approx\frac{1}{g}\big(|\partial\bm{n}/\partial t|^{2}-v_{s}^{2}|\partial\bm{n}/\partial x|^{2}\big).\label{eq:short_range_L}
\end{equation}
Here $g$ is an effective ($\alpha$- and short-distance-cutoff-dependent)
coupling strength, and the spin-wave velocity $v_{s}$ is also $\alpha$-dependent.
This model is gapped and disordered for all $g$ \cite{haldane_continuum_1983}.

To investigate ground state properties of Eq.\,(\ref{eq:short_range_L}),
we can remove the constraint $\bm{n}^{2}(x)=1$, while phenomenologically
introducing a mass gap $\Delta_{\alpha}$ and a renormalized spin-wave
velocity $v_{\alpha}$ (the parameters $\Delta_{\alpha}^{\prime}$
and $v_{\alpha}^{\prime}$ will be used to describe the Lagrangian
for $H_{\alpha}^{\prime}$) \cite{sorensen_equal-time_1994,sorensen_sk_1994}.
Transforming to momentum space, we thereby arrive at a free-field
Lagrangian density 
\begin{equation}
\mathcal{L}(q)\propto|\partial\bm{n}/\partial t|^{2}-(\Delta_{\alpha}^{2}+v_{\alpha}^{2}q^{2})|\bm{n}(q)|^{2}.\label{eq:L}
\end{equation}
This Lagrangian leads to ground-state correlations $\mathcal{C}_{ij}=\langle S_{i}^{z}S_{j}^{z}\rangle_{0}$
{[}where $\langle\cdots\rangle_{m}$ denotes the expectation value
in the state $|m\rangle$ defined in Fig.\,\ref{fig1}(a){]} that
decays as 
\begin{equation}
\mathcal{C}_{ij}\propto(-1)^{r}\int\frac{e^{iqr}dq}{\sqrt{\Delta_{\alpha}^{2}+v_{\alpha}^{2}q^{2}}}\propto(-1)^{r}K_{0}(r/\xi_{\alpha}).\label{eq:bulkcorr}
\end{equation}
Here $\xi_{\alpha}\equiv v_{\alpha}/\Delta_{\alpha}$ (or $\xi_{\alpha}^{\prime}\equiv v_{\alpha}^{\prime}/\Delta_{\alpha}^{\prime}$
for $H_{\alpha}^{\prime}$) defines the correlation length, and $K_{0}(x)$
is a modified Bessel function, which behaves as $K_{0}(x)\sim\exp(-x)/\sqrt{x}$
for large $x$.

For $\alpha<3$, the non-analytic $|q|^{\alpha-1}$ term in $H_{\alpha}^{\prime}$
dominates the dispersion of $\bm{n}(q)$ at small $q$, and Eqs.\,(\ref{eq:short_range_L}-\ref{eq:L})
are not valid. To analyze this case, we write down the renormalization
group (RG) flow equation for the coupling strength $g$ under the
scaling transformation $x\rightarrow xe^{-l}$ to one-loop order \cite{fradkin_field_2013,dutta_phase_2001},
\begin{equation}
\frac{dg}{dl}=\frac{\alpha-3}{2}g+\frac{g^{2}}{4\pi}.
\end{equation}
For $\alpha<3$, an unstable fixed point appears at $g^{\ast}=2\pi(3-\alpha)$,
and for a bare coupling $g<g^{\ast}$ the RG flow is towards a weak-coupling
ordered state at $g=0$ \cite{fradkin_field_2013}. The bare coupling,
and therefore the value of $\alpha$ at which this phase transition
occurs, is difficult to determine a priori. But we nevertheless expect
(and confirm numerically) that for $\alpha<\alpha_{c}$, with $2<\alpha_{c}<3$,
the gap will close as the system spontaneously breaks the continuous
SU(2) symmetry of $H_{\alpha}^{\prime}$ \cite{maghrebi_continuous_2015,gong1}.

\begin{figure}
\includegraphics[width=0.48\columnwidth]{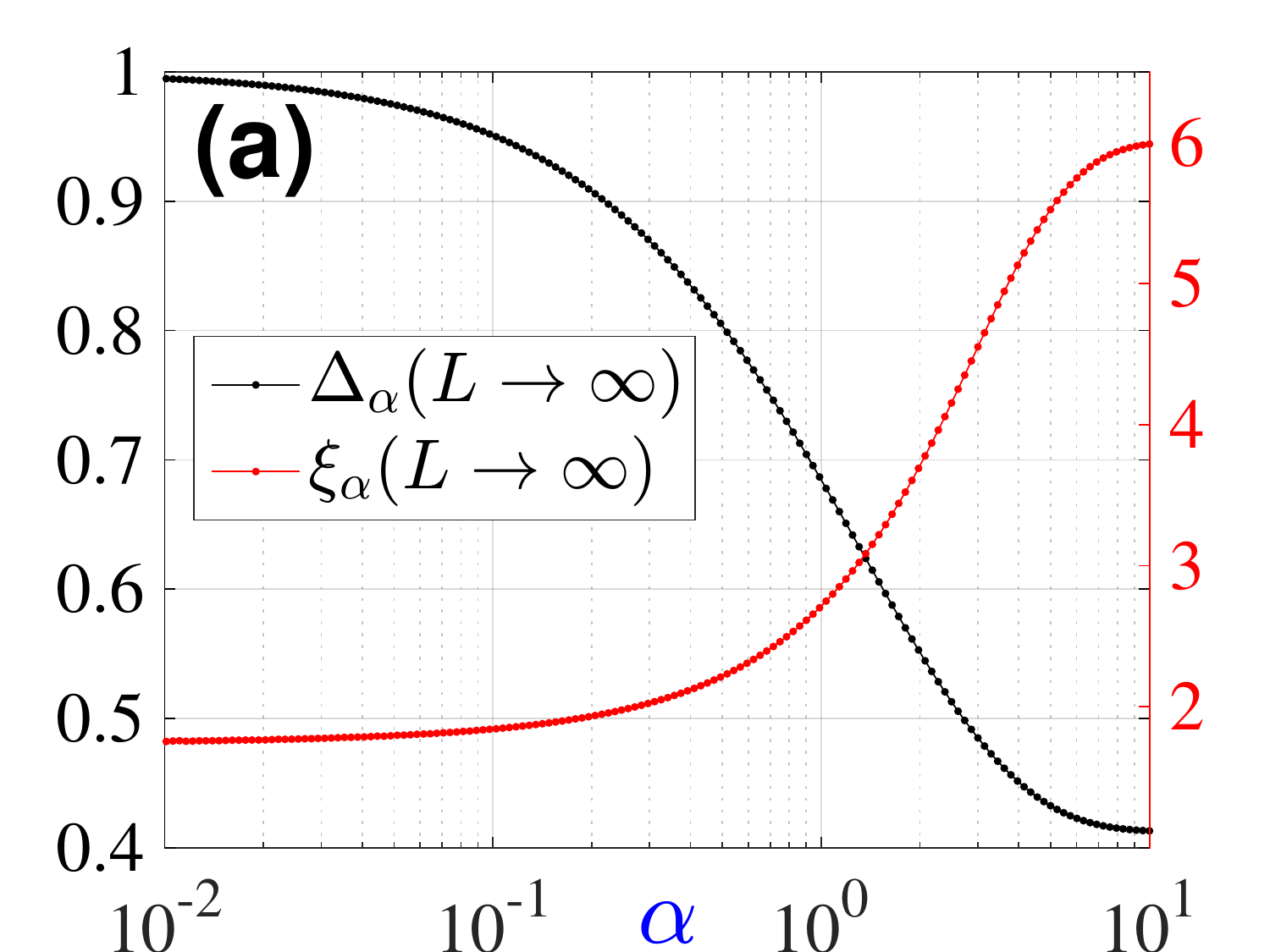} \includegraphics[width=0.48\columnwidth]{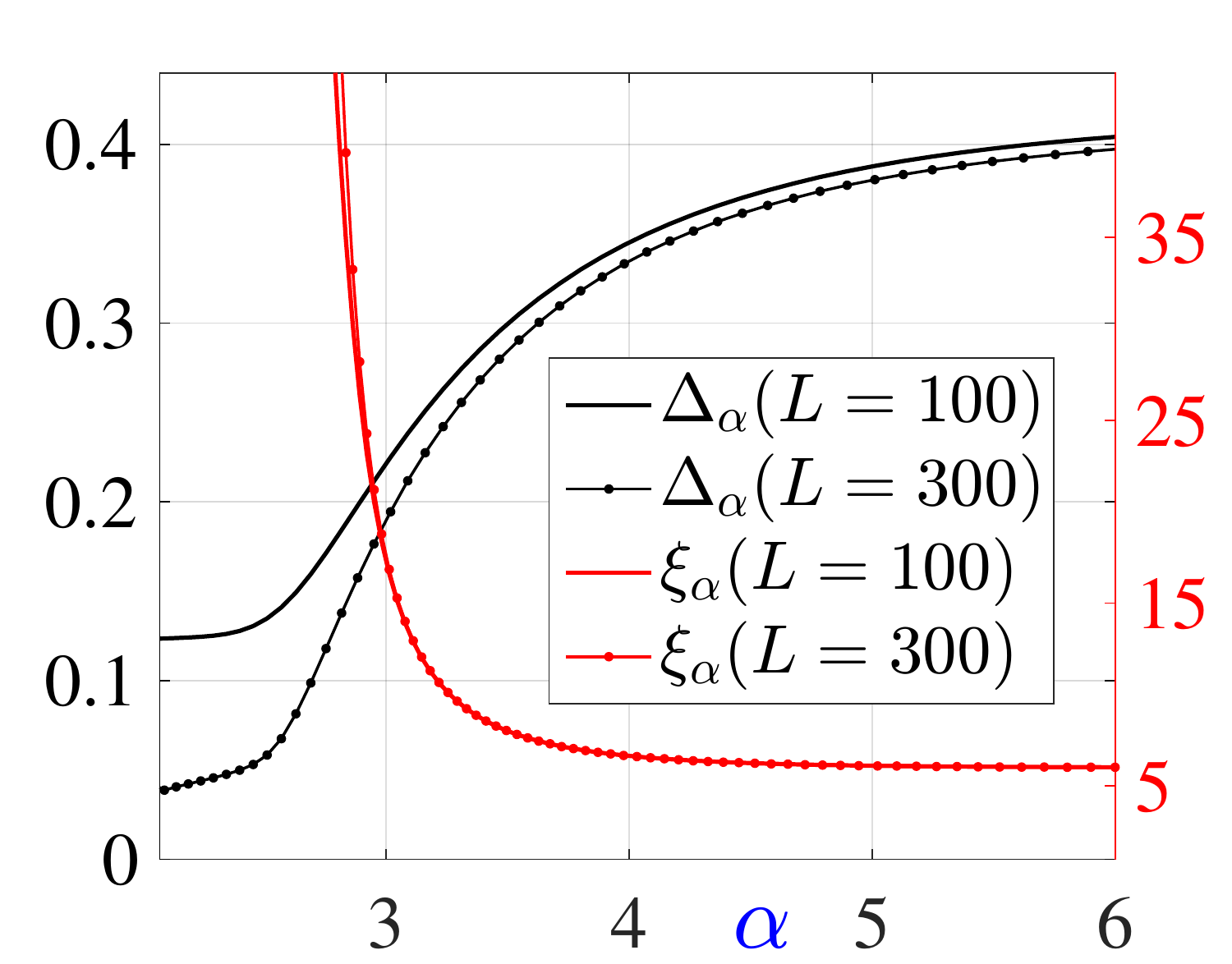}

\smallskip{}

\includegraphics[width=0.39\columnwidth]{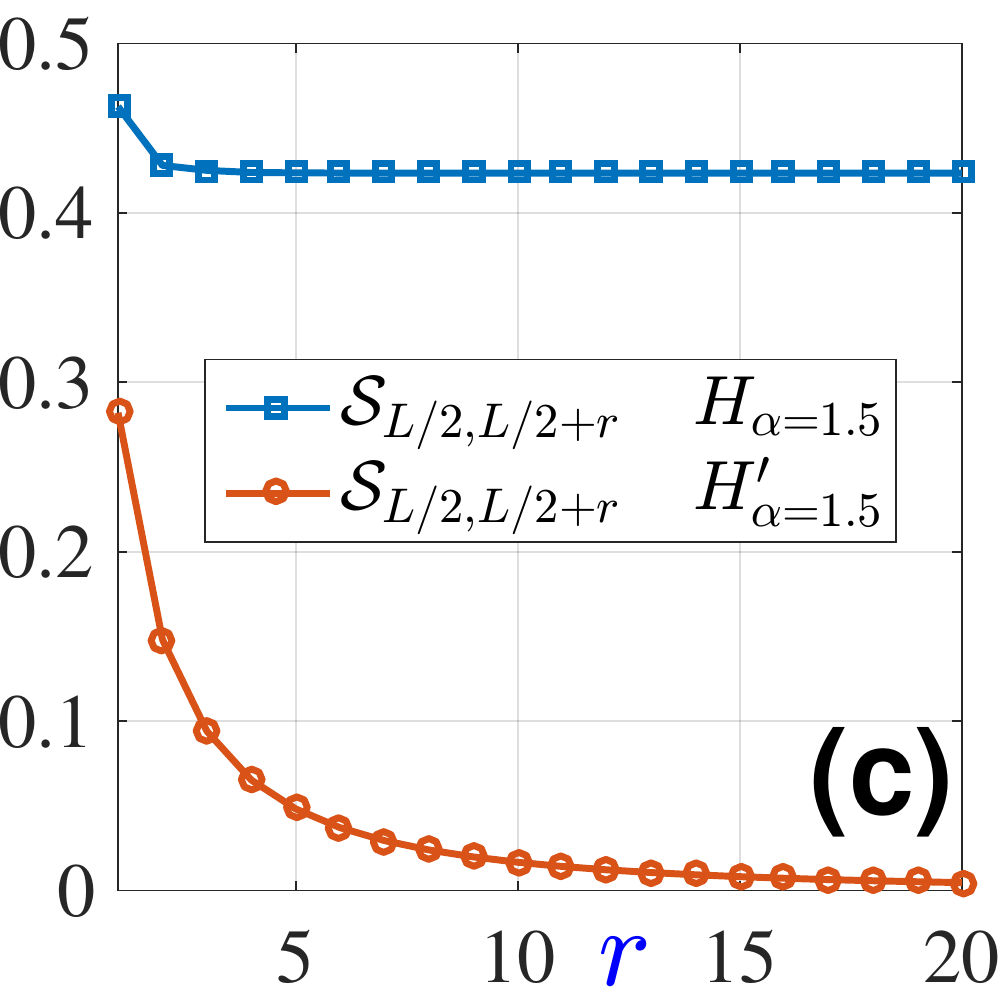} \includegraphics[width=0.58\columnwidth]{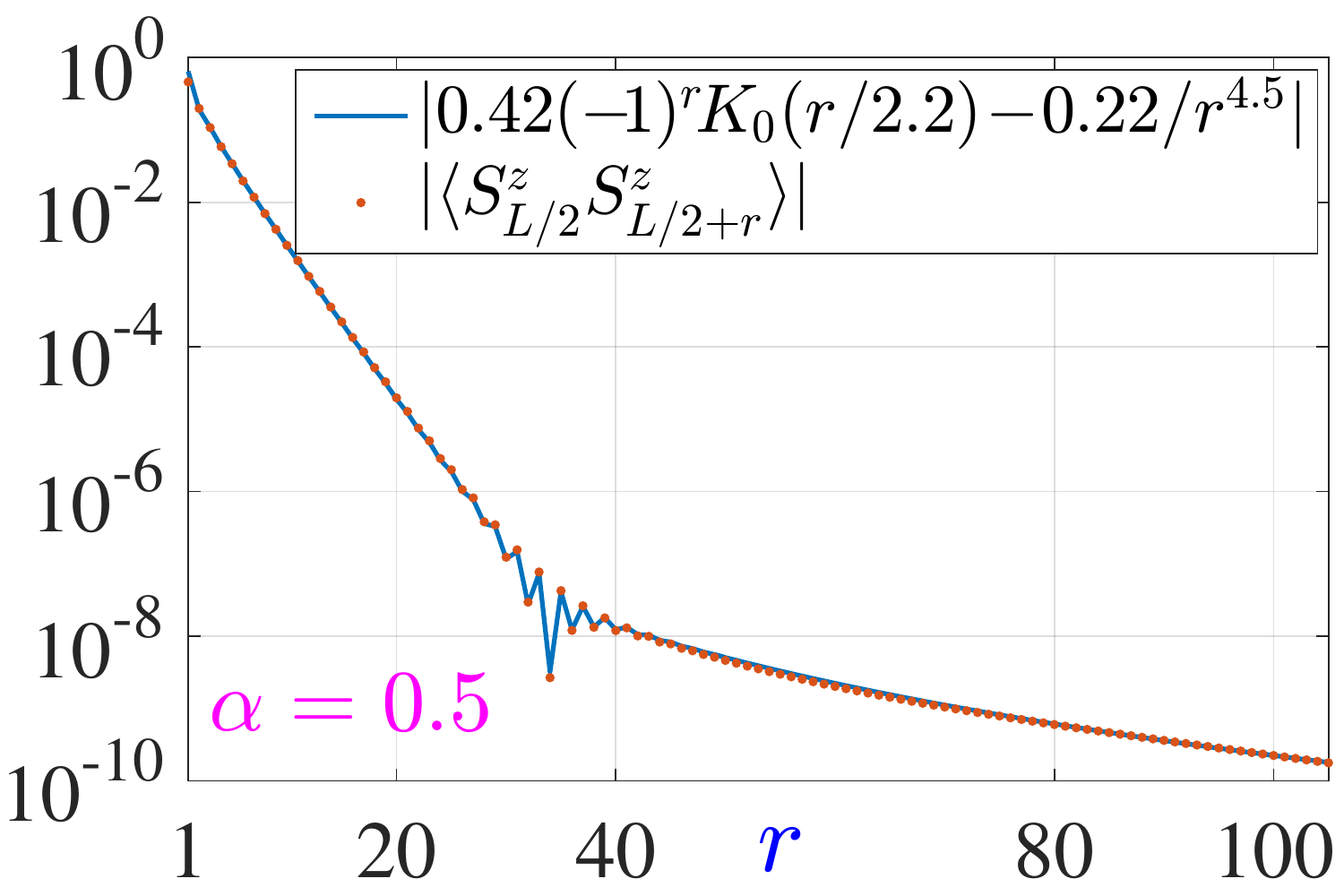}

\caption{(a) Bulk gap $\Delta_{\alpha}$ and ground state correlation length
$\xi_{\alpha}$ in the $L\rightarrow\infty$ limit, obtained by finite-size
scaling for $200\le L\le500$. (b) Bulk gap $\Delta_{\alpha}^{\prime}$
and $\xi_{\alpha}^{\prime}$ with $L=100$ and $L=300$. (c) Ground-state
string-ordered correlation function $\mathcal{S}_{ij}$ for $H_{\alpha}$
and $H_{\alpha}^{\prime}$ with $\alpha=1.5$ and $L=300$. For various
$\alpha$ and $200\le L\le500$, we consistently find that $\mathcal{S}_{ij}$
quickly saturates to a finite value for $H_{\alpha}$ at all $\alpha>0$,
but vanishes at large distance for $H_{\alpha}^{\prime}$ at $\alpha\lesssim3$.
(d) Ground-state spin-spin correlation $\mathcal{C}_{ij}$ for $\alpha=0.5$
and $L=500$. This choice of $\alpha=0.5$ is arbitrary, but assists
in a clear presentation of the coexisting exponential and $1/r^{\alpha+4}$
power-law decays.\label{fig2}}
\end{figure}

\emph{Comparison with numerics.—}Using finite-size MPS calculations,
we have obtained the bulk excitation gap $E_{2}-E_{1}$ and the correlation
length {[}fitted using Eq.\,(\ref{eq:bulkcorr}){]} for both $H_{\alpha}$
and $H_{\alpha}^{\prime}$. As shown in Fig.\,\ref{fig2}(a-b), we
see consistent results with the field theory predictions. For $H_{\alpha}$,
the gap remains open for all $\alpha>0$, and the correlation length
decreases together with $\alpha$ due to both an increase of the bulk
gap, and a decrease of the spin-wave velocity (as a result of a weakened
Néel order for longer-range interactions). To the contrary, for $H_{\alpha}^{\prime}$,
the gap decreases quickly as the interactions become longer ranged,
and the correlation length diverges when $\alpha$ decreases to around
$3$, suggesting the disappearance of the topological phase at $\alpha\lesssim3$
\cite{ft4}. Calculation of the string-ordered correlation $\mathcal{S}_{ij}\equiv\langle S_{i}^{z}S_{j}^{z}\prod_{i<k<j}(-1)^{S_{k}^{z}}\rangle_{0}$
of both $H_{\alpha}$ and $H_{\alpha}^{\prime}$ at $\alpha=1.5$
{[}Fig.\,\ref{fig2}(c){]} provides further evidence that the topological
phase survives for $H_{\alpha}$, but not for $H_{\alpha}^{\prime}$,
for $0<\alpha\lesssim3$.

We now analyze the effects of terms beyond leading order in $q$ that
have been ignored in our field theory treatment. Including the higher-order
analytic terms, such as the $O(q^{4})$ term, will result in negligible
corrections to the correlation functions that decay in distance faster
than Eq.\,(\ref{eq:bulkcorr}) \cite{sorensen_equal-time_1994}.
However, even for $\alpha>3$, inclusion of the non-analytic $O(|q|^{\alpha-1})$
term will add a power-law tail to the correlation functions, which
will dominate over Eq.\,(\ref{eq:bulkcorr}) at long distance. In
the supplemental material, we show by a more involved field-theory
calculation that, for $H_{\alpha}$, $\mathcal{C}_{ij}$ decays as
$1/r^{\alpha+4}$ at large $r$. Our MPS calculations using $L=500$
spins {[}Fig.\,\ref{fig2}(d){]} show remarkable agreement with the
field theory predictions, even capturing the oscillations in $|\mathcal{C}_{ij}|$
occurring at intermediate distance where the short-range and long-range
contributions to the correlation functions are of comparable magnitude
and interfere. A power-law tail in $\mathcal{C}_{ij}$ should also
exist for $H_{\alpha}^{\prime}$, but the increased correlation length
prevents us from observing its existence clearly for $\alpha>3$.

\begin{figure}[t]
\includegraphics[width=0.49\columnwidth]{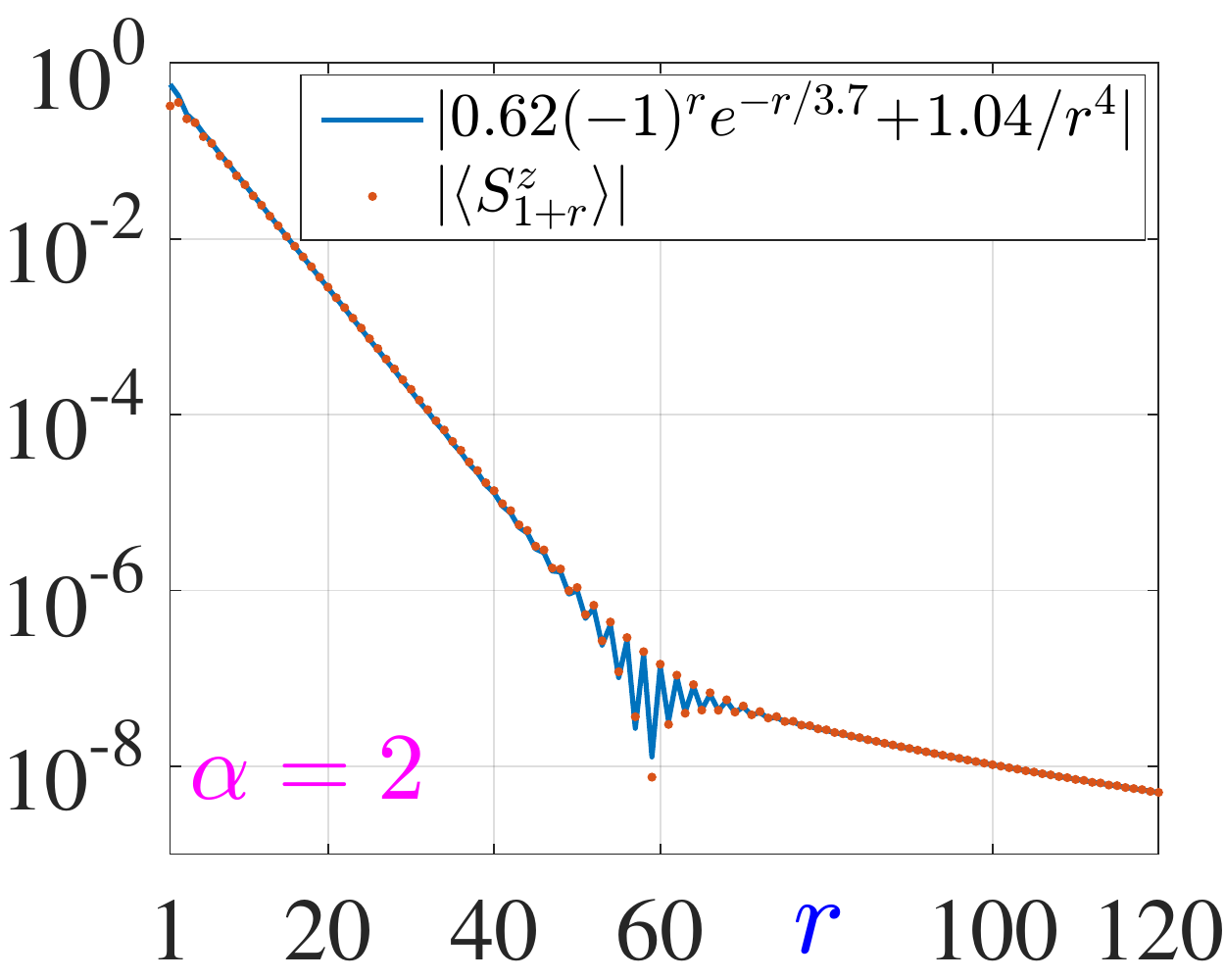}\includegraphics[width=0.49\columnwidth]{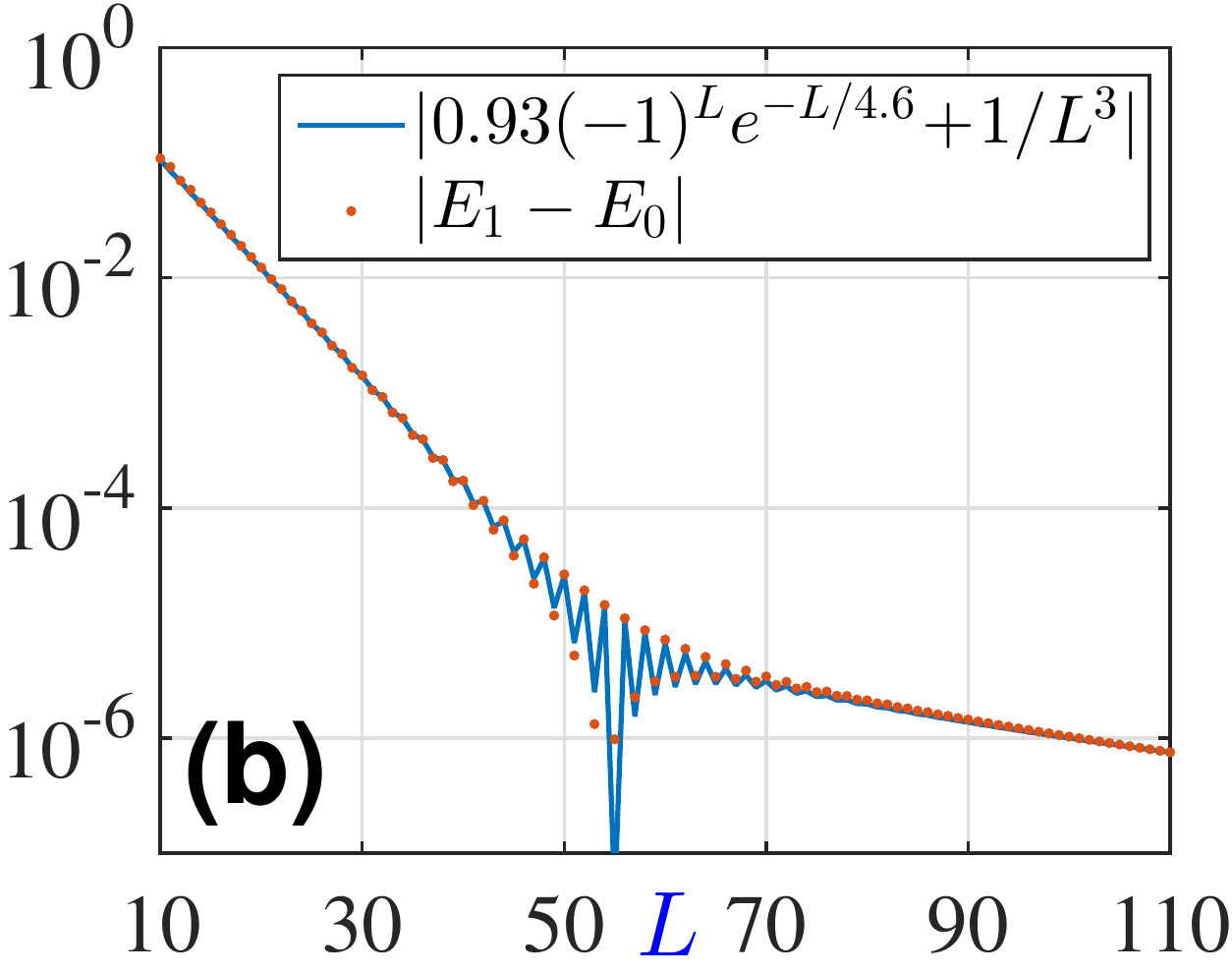}

\includegraphics[width=0.59\columnwidth]{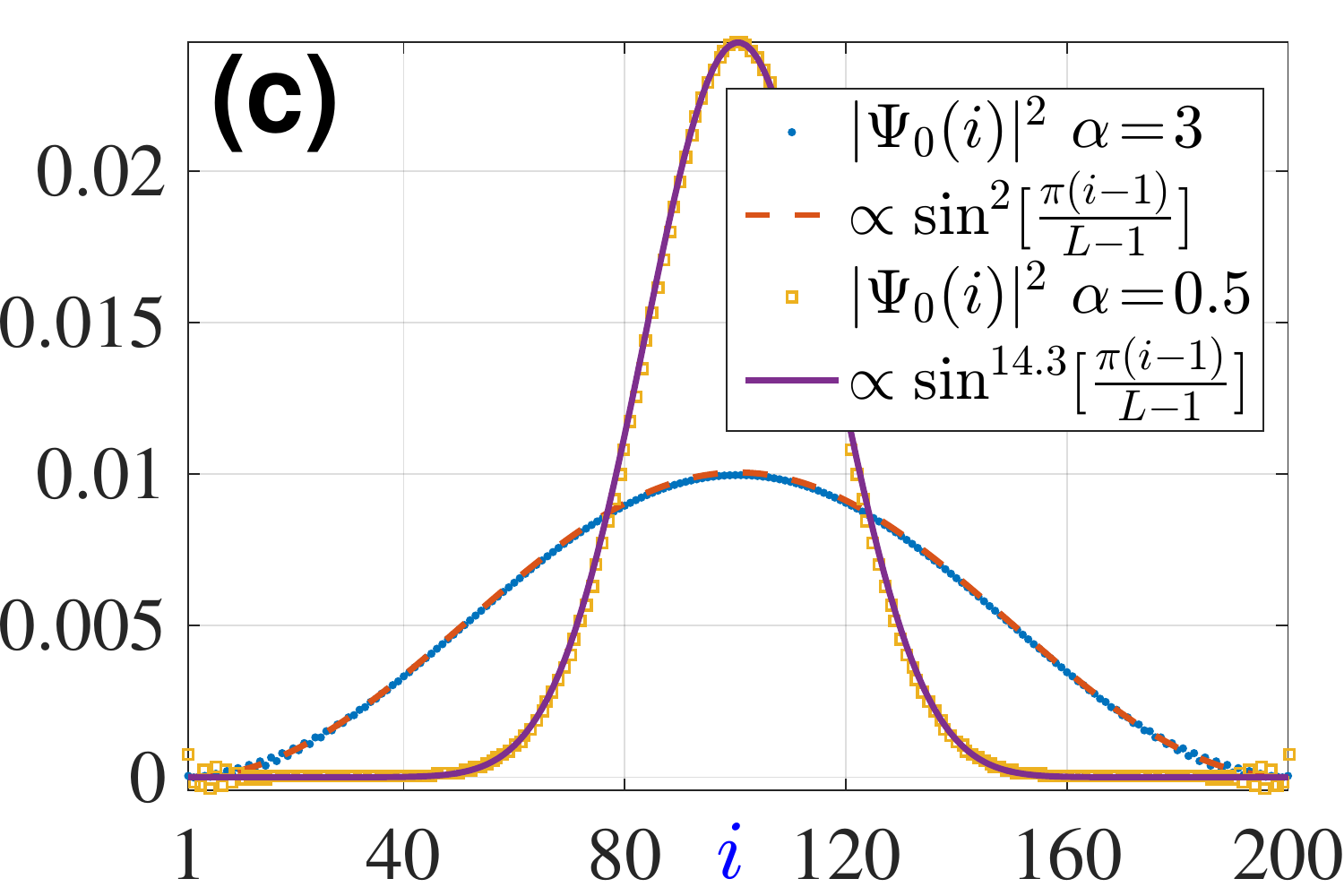} \includegraphics[width=0.4\columnwidth]{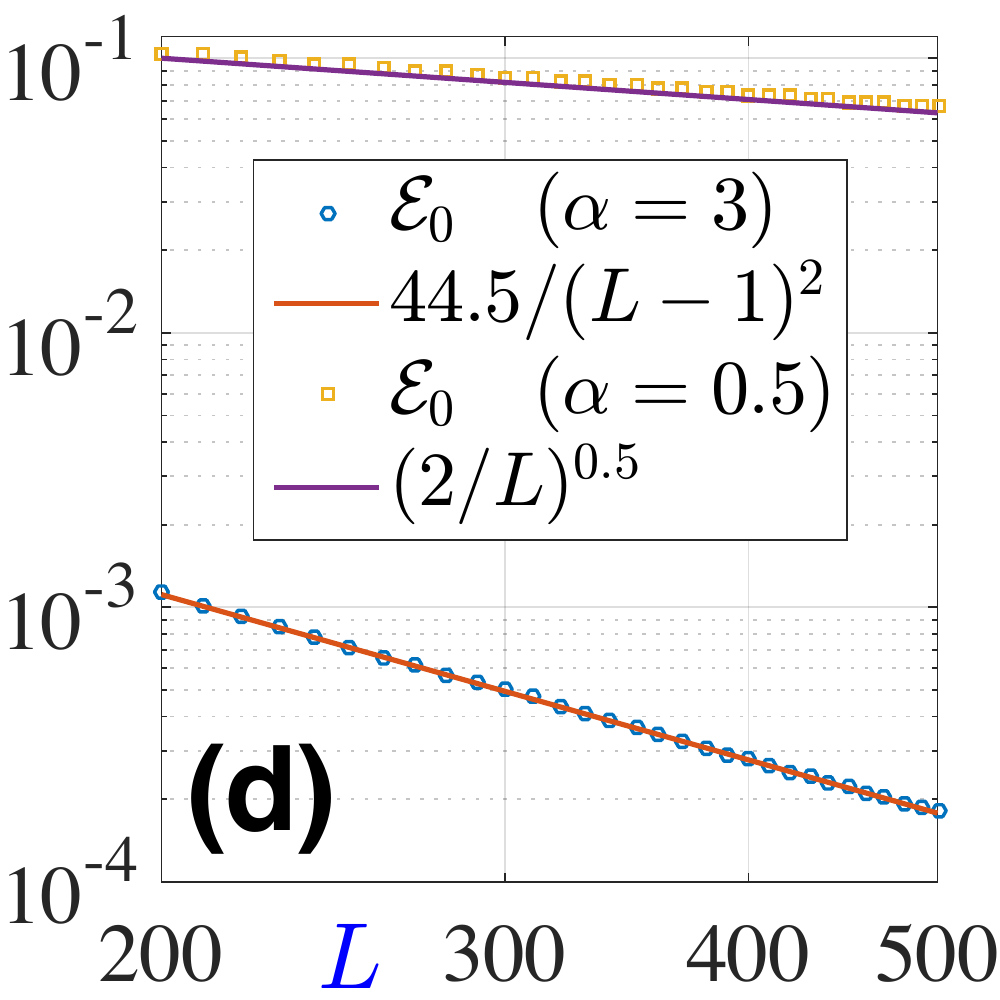}

\caption{(a) Distribution of an edge excitation in state $|1\rangle$ for $L=500$
and $\alpha=2$. (b) Edge gap $|E_{1}-E_{0}|$ as a function of the
chain size $L$ for $\alpha=3$.\label{fig3} (c) Lowest energy magnon
probability density distribution for $L=200$ and $\alpha=3.0,0.5$.
(d) The finite-size correction to the lowest magnon excitation energy
{[}see Eq.\,(\ref{eq:MagSch}){]}. For $\alpha=3$, we obtain $v_{\alpha}=2.18$
and $v_{\alpha}/\Delta_{\alpha}\approx4.51$, in good agreement with
the $\xi_{\alpha}\approx4.55$ obtained in Fig.\,\ref{fig2}.}
\end{figure}

\emph{Edge excited states.—}We expect the influence of long-range
interactions on the edge- and bulk-excited states to be strong at
small $\alpha$; because the topological phase of $H_{\alpha}^{\prime}$
does not survive for $\alpha\lesssim3$, we will focus on $H_{\alpha}$
from now on. Edges can be introduced into the field theory by replacing
the two end spin-1's with spin-1/2's, represented by $\bm{\tau}_{L}$
($\bm{\tau}_{R}$) for the left (right) edge, resulting in an edge-bulk
coupling Hamiltonian $H_{c}=\sum_{i=2}^{L-1}\bm{S}_{i}\cdot\big(\bm{\tau}_{L}/(i-1)^{\alpha}+\bm{\tau}_{R}/(L-i)^{\alpha}\big)$
\cite{sorensen_equal-time_1994}. For the edge excited state $|1\rangle$
{[}Fig.\,\ref{fig1}(a){]}, $\bm{\tau}_{L,R}$ are polarized in the
$+z$ direction, and we expect $\langle S_{i}^{z}\rangle$ to decay
away from the ends. Solving the free theory defined by Eq.\,(\ref{eq:L})
and treating $H_{c}$ using standard first-order perturbation theory
\cite{sorensen_equal-time_1994}, we find that $\langle n^{z}(x)\rangle_{1}\propto\int dq(\exp[iq(L-x)]-\exp[iq(x-1)])/(\Delta_{\alpha}^{2}+v_{\alpha}^{2}q^{2})\propto\exp[-(L-x)/\xi_{\alpha}]-\exp[-(x-1)/\xi_{\alpha}]$
for even $L$. In addition, $\langle l^{z}(x)\rangle_{1}$ contributes
a power-law correction $1/(x-1)^{\alpha+2}+1/(L-x)^{\alpha+2}$ for
$x$ far away from both ends \cite{supp}. Our numerical calculation
of $\langle S^{z}(x)\rangle_{1}$, shown in Fig.\,\ref{fig3}(a),
agrees well with a sum of these two contributions, clearly exhibiting
an exponential followed by $1/r^{\alpha+2}$ decay.

The edge gap $|E_{1}-E_{0}|$ can be obtained by using a path integral
to integrate out the $\bm{n}$ field \cite{sorensen_equal-time_1994},
resulting in an effective edge-edge Hamiltonian $\propto(-1)^{L}\exp(-L/\xi_{\alpha})\bm{\tau}_{L}\cdot\bm{\tau}_{R}$.
This scaling is confirmed, at relatively small $L$, by the numerical
results in Fig.\,\ref{fig3}(b). However, the numerics also reveal
that at large $L$ the edge gap receives a long-range correction given
by $1/L^{\alpha}$. This remarkably simple result, including the unity
prefactor, can be understood as follows. The edge-excited states behave
differently from the bulk-excited states due to \emph{correlations}
between the orientations of $\bm{\tau}_{1}$ and $\bm{\tau}_{2}$,
and therefore $\langle\bm{S}_{i}\cdot\bm{S}_{j}\rangle_{1}-\langle\bm{S}_{i}\cdot\bm{S}_{j}\rangle_{0}$
is very small unless $i$ and $j$ are very close to $0$ and $L$,
respectively. Thus we have $E_{1}-E_{0}\approx L^{-\alpha}\sum_{i<j}(\langle\bm{S}_{i}\cdot\bm{S}_{j}\rangle_{1}-\langle\bm{S}_{i}\cdot\bm{S}_{j}\rangle_{0})=1/L^{\alpha}$,
where the last equality is a sum-rule following from the total spin
of the ground ($S=0$) and edge excited ($S=1$) states.

\emph{Bulk excited states.—}As in the short-range Haldane chain, the
elementary bulk excitations of $H_{\alpha}$ are spin-1 magnons \cite{white_numerical_1993,sorensen_equal-time_1994,sorensen_large-scale_1993}.
Physically, the magnon represents fluctuations in the field $\bm{n}$,
and, from Eq.\,(\ref{eq:L}), these fluctuations have a dispersion
relation $\epsilon_{\alpha}(q)=\sqrt{\Delta_{\alpha}^{2}+(v_{\alpha}q)^{2}}\approx\Delta_{\alpha}+q^{2}v_{\alpha}^{2}/(2\Delta_{\alpha})$
(valid at small $q$). The lowest-energy magnon wave-function $\Psi_{0}(x)$
can be extracted from the numerics using the relation $|\Psi_{0}(i)|^{2}\approx|\langle S_{i}^{z}\rangle_{2}-\langle S_{i}^{z}\rangle_{1}|$.
The presence of long-range interactions gives the magnon an additional
potential energy due to the edge-bulk coupling Hamiltonian $H_{c}$,
and $\Psi(x)$ can be approximately described by the following Schrödinger
equation (with Dirichlet boundary condition at $x=1,L$) 
\begin{equation}
\!\!\!\frac{v_{\alpha}^{2}}{2\Delta_{\alpha}}\!\frac{\partial^{2}\Psi(x)}{\partial x^{2}}\!+\!\frac{1}{2}\!\left[\!\frac{1}{(x-1)^{\alpha}}\!+\!\frac{1}{(L-x)^{\alpha}}\!\right]\!\Psi(x)\!=\!\mathcal{E}\Psi(x).\label{eq:MagSch}
\end{equation}
The kinetic (potential) energy always scales as $1/L^{2}$ ($1/L^{\alpha}$);
therefore, for $\alpha>2$ and large $L$, the potential energy can
be ignored. The ground-state energy $\mathcal{E}_{0}\approx v_{\alpha}^{2}\pi^{2}/(2\Delta_{\alpha}L^{2})$
and probability density $|\Psi_{0}(x)|^{2}\approx(2/L)\sin^{2}(\pi x/L)$
are then identical to those of a particle in a box, as confirmed numerically
in Fig.\,\ref{fig3}(c-d). The relation $E_{2}-E_{1}\approx\Delta_{\alpha}+v_{\alpha}^{2}\pi^{2}/(2\Delta_{\alpha}L^{2})$
allows us to obtain both $v_{\alpha}$ and $\Delta_{\alpha}$ through
finite-size scaling {[}Fig.\,\ref{fig2}(b){]}, and we confirm that
the correlation length determined by $\xi_{\alpha}=v_{\alpha}/\Delta_{\alpha}$
agrees with that obtained by fitting $\mathcal{C}_{ij}$ using Eq.\,(\ref{eq:bulkcorr}).
For $\alpha<2$, the potential energy dominates the kinetic energy
for large $L$, and the potential can be approximated as harmonic
around $x=L/2$. Thus $|\Psi_{0}(x)|^{2}$ resembles a Gaussian {[}Fig.\,\ref{fig3}(c){]},
and a simple scaling analysis predicts a width $\gamma\propto L^{1-\alpha/2}$.
In the large-$L$ limit, $|\Psi_{0}(x)|^{2}$ become sharply peaked
at $x=L/2$ and, from Eq.\,(\ref{eq:MagSch}), we expect the bulk
gap to scale as $\Delta_{\alpha}+(2/L)^{\alpha}$, which is clearly
observed in Fig.\,\ref{fig3}(d). Since $E_{2}-E_{1}=2$ when $\alpha=0$,
it follows that $\Delta_{\alpha\rightarrow0}=1$, consistent with
Fig.\,\ref{fig2}(a).

\emph{Outlook.—}The stability of the topological Haldane phase to
$1/r^{\alpha}$ interactions for all $\alpha>0$ is favorable for
trapped-ion based experiments, as stronger couplings can be achieved
for smaller $\alpha$ \cite{richerme_non-local_2014,jurcevic_quasiparticle_2014}.
Moreover, because the correlation length \emph{shrinks} for longer-range
interactions, a relatively small number of ions will suffice to suppress
finite-size effects. Probing the topological phase by measuring both
$\mathcal{C}_{ij}$ and $\mathcal{S}_{ij}$ with single-site resolution
is nearly impossible in typical condensed-matter systems, but is quite
straightforward in ion traps \cite{cohen_simulating_2015}. Based
on the generality of our field theory analysis, we speculate that
for generic lattice models, the tails in the power-law interactions
can possibly destroy the topological phase only when long-range interactions
are unfrustrated and $\alpha<D+2$. Experimentally, unfrustrated long-range
interactions can be easily implemented by generating a $1/r^{\alpha}$
ferromagnetic interaction \cite{gong1}. We hope that our work can
serve as a springboard for future studies on how distinct topological
phases behave in the presence of long-range interactions.
\begin{acknowledgments}
We thank C.\ Monroe, G.\ Pupillo, A.\ Turner, J. Sau, M.\,Hafezi,
J.\,Pixley, P.\ Richerme, C.\ Senko, P.\ Hess, B.\ Neyenhuis,
A.\ Lee, J.\ Smith, A.\ M.\ Rey, S.\ Manmana, and K.\ Hazzard
for helpful discussions. This work was supported by the AFOSR, NSF
PIF, the ARO, NSF PFC at the JQI, the ARL, and the AFOSR MURI. M.
F.-F. and M.\,L. W. thank the NRC for support. 
\end{acknowledgments}

 \bibliographystyle{apsrev4-1}
\bibliography{sub}

\begin{thebibliography}{74}
\bibitem [{ft1()}]{ft1}%
  \BibitemOpen
  \href@noop {} {\bibinfo  {journal} {Note that for symmetry protected
  topological phases, we assume that perturbations (either local or
  long-ranged) do not break the symmetries protecting the topological
  phases\kern-0.25em}\ }\BibitemShut {NoStop}%
\bibitem [{ft2()}]{ft2}%
  \BibitemOpen
  \href@noop {} {\bibinfo  {journal} {Since we are only concerned about the
  small-q components of both fields $\bm{n}$ and $\bm{l}$ at low energy, the
  cross terms between $\bm{n}$ and $\bm{l}$ in $H_{\alpha}$ and
  $H_{\alpha}^{\prime}$ can be ignored since they involve $\bm{n}(q)$ or
  $\bm{l}(q)$ near $q=\pi$\kern-0.25em}\ }\BibitemShut {NoStop}%
\bibitem [{ft4()}]{ft4}%
  \BibitemOpen
  \href@noop {} {\bibinfo  {journal} {The divergent correlation length makes it
  difficult to accurately obtain the critical $\alpha$ from the system sizes
  accessed here, but the existence of a critical $\alpha_{c}$ between 2 and 3
  is further supported by infinite-size MPS calculations performed in
  \cite{gong1}\kern-0.25em}\ }\BibitemShut {NoStop}%
\end{thebibliography}

\end{document}